# Si-based GeSn photodetectors towards mid-infrared imaging applications

Huong Tran,[1,2,a)] Thach Pham,[1,2,a)] Joe Margetis,[3] Yiyin Zhou,[1,2,4] Wei Dou,[1] Perry C. Grant,[1,2,4] Joshua M. Grant,[1,4] Sattar Alkabi,[1,4] Greg Sun,[5] Richard A. Soref,[5] John Tolle,[3] Yong-Hang Zhang,[6] Wei Du,[7] Baohua Li,[2] Mansour Mortazavi,[8] and Shui-Qing Yu[1, *]

[1]*Department of Electrical Engineering, University of Arkansas, Fayetteville, Arkansas 72701, USA*

[2]*Arktonics, LLC, 1339 South Pinnacle Drive, Fayetteville, Arkansas 72701, USA*

[3]*ASM, 3440 East University Drive, Phoenix, Arizona 85034, USA*

[4]*Microelectronics-Photonics Program, University of Arkansas, Fayetteville, Arkansas 72701, USA*

[5]*Department of Engineering, University of Massachusetts Boston, Boston, Massachusetts 02125, USA*

[6]*Department of Electrical Engineering, Arizona State University, Tempe, Arizona 85281, USA*

[7]*Department of Electrical Engineering and Physics, Wilkes University, Wilkes-Barre, Pennsylvania 18766, USA*

[8]*Department of Chemistry and Physics, University of Arkansas at Pine Bluff, Pine Bluff, Arkansas 71601, USA*

## Abstract

This paper reports a comprehensive study of Si-based GeSn mid-infrared photodetectors, which includes: i) the demonstration of a set of photoconductors with Sn compositions ranging from 10.5% to 22.3%, showing the cut-off wavelength has been extended to 3.65 µm. The measured maximum D* of $1.1 \times 10^{10}$ cm·Hz$^{1/2}$·W$^{-1}$ is comparable to that of commercial extended-InGaAs detectors; ii) the development of surface passivation technique on photodiode based on in-depth analysis of dark current mechanism, effectively reducing the dark current. Moreover, mid-infrared images were obtained using GeSn photodetectors, showing the comparable image quality with that acquired by using commercial PbSe detectors.

---

[a)] Huong Tran and Thach Pham contributed equally to this work
[*] Corresponding author email: syu@uark.edu

Low-cost near- and short-wave infrared (NIR, SWIR) photodetectors have great potential for night-vision applications, both in defense and civilian systems including video surveillance security cameras, future automotive night vision, integrated night vision in smart phones, and in other mobile/wearable electronics. The current market-dominating SWIR photodetectors mainly rely on expensive group III-V (such as InGaAs and InSb) and group II-VI (such as HgCdTe) materials. For the past decades, many efforts have been made towards the hybrid integration of III-Vs or II-VIs on a Si substrate. It is highly desirable to develop an alternative material featuring lower-cost and high-performance for SWIR photodetectors.[1-4] Recently, GeSn techniques have drawn much attention as the success of developing GeSn alloy has opened an avenue for a totally new generation of infrared detectors. Research activities in GeSn-based detectors have experienced a strong increase in the past a few years. Developing high-performance GeSn detectors has been enabled by the following progresses: i) broad operation wavelength coverage up to 12 μm by increasing the Sn compositions;[5-8] ii) true direct bandgap allowing for enhanced band-to-band light absorption;[5-6] iii) a complete Si complementary metal-oxide-semiconductor (CMOS) compatible process for material growth; iv) a viable low-cost solution for large-scale focal plane arrays (FPAs) monolithically integrated on Si; v) a readily available growth technique using industry standard reactors to reach device-level material quality;[1,4,7-11] and vi) a GeSn-based 320×256 FPA imaging sensor with spectra response in the 1.6-1.9 μm range was successfully demonstrated in 2016.[12]

In the last a few years, many GeSn-based photodetectors including photoconductors and photodiodes have been reported with dramatically improved performance. Moreover, the advanced structures such as double-heterostructure structure (DHS) and quantum well (QW) structure were employed to enhance the carrier confinement. A brief review of GeSn detector development is shown in Table I.[13-40] To fully understand the potential of GeSn detectors, a comprehensive study regarding Sn-compositional device performance, particular with higher Sn composition (higher than 10%), is highly desirable.

Table I: A brief review of GeSn photodetector development

| Year | Sn% | Structure | Responsivity @1.55 μm (A/W) | Cut-off (μm) | Dark current | Reference |
|------|-----|-----------|------------------------------|--------------|--------------|-----------|
| 2009 | 2% | PIN | 0.05 | 1.75 | 1 A/cm$^2$ | [13] |
| 2011 | 0.5% | PIN | 0.1 | 1.65 | 10 A/cm$^2$ | [14] |
| 2011 | 2% | PIN | 0.12 | 1.75 | 1.0 A/cm$^2$ | [15] |
| 2011 | 3% | PIN | 0.23, 0.12 @1.64 μm | 1.8 | 1.8 A/cm$^2$ | [16] |
| 2013 | 3.9% | PIN | 0.27, 0.165 @1.6 μm | 1.8 | 400 mA/cm$^2$ | [17] |
| 2012 | 4% | PIN | 0.2, 0.17@1.6 μm | 1.95 | 100 mA/cm$^2$ | [18] |
| 2012 | 9% | QW/PC | 1.0, 0.1 @ 2.2 μm | 2.3 | 0.5 mA | [19] |
| 2013 | 3.6% | PIN | 0.71 @ 1.8μm | 2.0 | 6.1 mA/cm$^2$ | [20] |
| 2013 | 8% | PN | 0.031 | 1.575 | 11 μA | [21] |
| 2013 | 9.8% | PN | Broadband photocurrent 200 μA | 2.0 | 100 μA | [22] |
| 2013 | 9% | PN | N. A. | No report | 1 A/cm$^2$ | [23] |
| 2014 | 1.8% | PIN | 0.18 | 1.6 | 50 mA/cm$^2$ | [24] |
| 2014 | 4.2% | PIN | 0.22 | N. A. | 890 mA/cm$^2$ | [25] |
| 2014 | 7% | QW/PIN | 0.13 | 1.91 | 100 mA/cm$^2$ | [26] |
| *2014* | *7%* | *PC* | *0.18 @ 10 V* | *2.1* | *37.5 mA/cm$^2$ @10V* | *[27]* |
| *2014* | *10%* | *PC* | *1.63 @ 50 V* | *2.4* | *82.5 mA/cm$^2$* | *[28]* |
| 2014 | 12% | PN | 0.081 | 2.0 @ 100 K | 50 mA | [29] |
| 2015 | 5% | PIN | 0.18, 0.06 @ 1.63 μm | 1.85 | 73 mA/cm$^2$ | [30] |
| *2015* | *10%* | *PC* | *2.85 @ 5V* | *2.4* | *7.7 A/cm$^2$* | *[31]* |
| 2016 | 2.5% | PIN | 0.3 | 1.9 | 1 mA/cm$^2$ | [32] |
| 2016 | 2.5% | PIN/Array | 0.45 @ 1.7 μm | 1.9 | 0.1 mA/cm$^2$ | [12] |
| 2016 | 6% | PIN | 0.35, 0.09 @ 2 μm | 1.985 | 3 A/cm$^2$ | [33], [34] |
| *2016* | *7%* | *PIN* | *0.3* | *2.2 μm* | *1A/cm$^2$* | *[35]* |
| *2016* | *10%* | *PIN* | *0.19* | *2.6 μm* | *4.3 A/cm$^2$* | *[35]* |
| 2017 | 2.8% | MQW/PIN | 62 mA/W | 1.635 | 59 mA/cm$^2$ | [36] |
| 2017 | 3% | QW/PIN | 80 mA/W | 1.65 | 4 mA/cm$^2$ | [37] |
| 2017 | 10% | MQW PIN | 0.02 @ 2 μm | N. A. | 31 mA/cm$^2$ | [38] |
| 2017 | 6.5% | HPT | 1.8, 0.043 @ 2 μm | 2.0 | 147 mA/cm$^2$ | [39] |
| *2018* | *11%* | *PIN* | *0.32 A/W @ 2 μm* | *2.65 μm* | *7.9 mA/cm$^2$* | *[40]* |

* PC stands for photoconductor; The results in bold and italic are reported by our group.

We previously reported GeSn photoconductors with Sn compositions up to 10%, featuring high peak responsivity of 2.85 A/W at a wavelength of 1.55 μm due to high photoconductive gain using interdigitated electrodes.[31]  Moreover, p-i-n photodiodes with Sn compositions ranging from 7% to 11% were investigated, showing the cut-off wavelength extended to 2.65 μm.[35]  These results provided a baseline for device performance at SWIR wavelength range.  In this paper, we report the demonstration of: i) a set of GeSn photoconductors with Sn compositions from 10.5% to 22.3% and cut-off wavelengths up to 3.65 μm.  The highest specific detectivity (D*) of $1.1×10^{10}$ cm·Hz$^{1/2}$·W$^{-1}$ was obtained, which is comparable with that ($\sim4×10^{10}$ cm·Hz$^{1/2}$·W$^{-1}$) of commercial extended-InGaAs detectors;  and ii) the passivation technique on GeSn photodiode was investigated.  Based on the in-depth current-voltage analysis, the significantly reduced dark current was achieved with passivated device.  In addition, we demonstrated the mid-infrared imaging using both photoconductors and photodiodes, showing comparable image quality with commercial PbSe detector.  Particularly, the images from photoconductor were taken without lock-in technique.

**Results and discussions**

**Material growth and characterization.**  The GeSn samples were grown using an industry standard ASM Epsilon® 2000 Plus reduced pressure chemical vapor deposition system (RPCVD) with commercially available precursors.  The material characterization results including layer thickness and Sn composition are summarized in Table II.  The samples are annotated from A to F according to the Sn compositions.

Table II Summary of GeSn samples

| Sample | GeSn layer | Thickness (nm) | Spontaneous relaxed gradient Sn% (by SIMS) |
|--------|-----------|----------------|---------------------------------------------|
| A | 1st layer | 180 | 8.8% – 10.5% |
|   | 2nd layer | 660 | 10.5% – 12.5% |
| B | 1st layer | 250 | 12.5% – 13.2% |
|   | 2nd layer | 670 | 13.2% – 15.9% |
| C | 1st layer | 165 | 9.2% – 10.0% |
|   | 2nd layer | 585 | 10.0% – 13.2% |
|   | 3rd layer | 254 | 13.2% – 15.7% |
| D | 1st layer | 310 | 11.6% – 12.2% |
|   | 2nd layer | 550 | 12.2% – 16.2% |
|   | 3rd layer | 260 | 16.2% – 17.9% |
| E | 1st layer | 450 | 11.5% – 15.3% |
|   | 2nd layer | 950 | 15.3% – 20.0% |
| F | 1st layer | 380 | 11.8% – 15.5% |
|   | 2nd layer | 830 | 15.5% – 22.3% |

Transmission electron microscopy (TEM) image for each sample shows clearly resolved layer structure. For GeSn, the distinct multi-layer feature was observed. Figure 1 (a) shows a typical TEM image of sample E. Two distinct layers that can be resolved including a 450-nm-thick defective bottom layer and a 950-nm-thick high-quality top layer. From the secondary ion mass spectrometry (SIMS) (not shown here), both layers exhibit a spontaneously Sn-enhanced gradient, from 11.5% to 15.3%, and from 15.3% to 20% for bottom and top layers, respectively. The increased Sn composition is mainly due to the ease of compressive strain, which facilitates the Sn incorporation. Although the bottom GeSn layer is defective, the formation of threading dislocation loops in the bottom layer prevents the defects from propagating into the top GeSn layer, resulting in high-quality top layer.[41] The detailed growth mechanism can be found elsewhere.[9, 10, 41] It is worth pointing out that based on our simulation, the current flow path is not only through the top high-quality GeSn layer, but also includes the bottom defective GeSn layer as well as the Ge buffer layer, as shown in Fig. 1 (b).

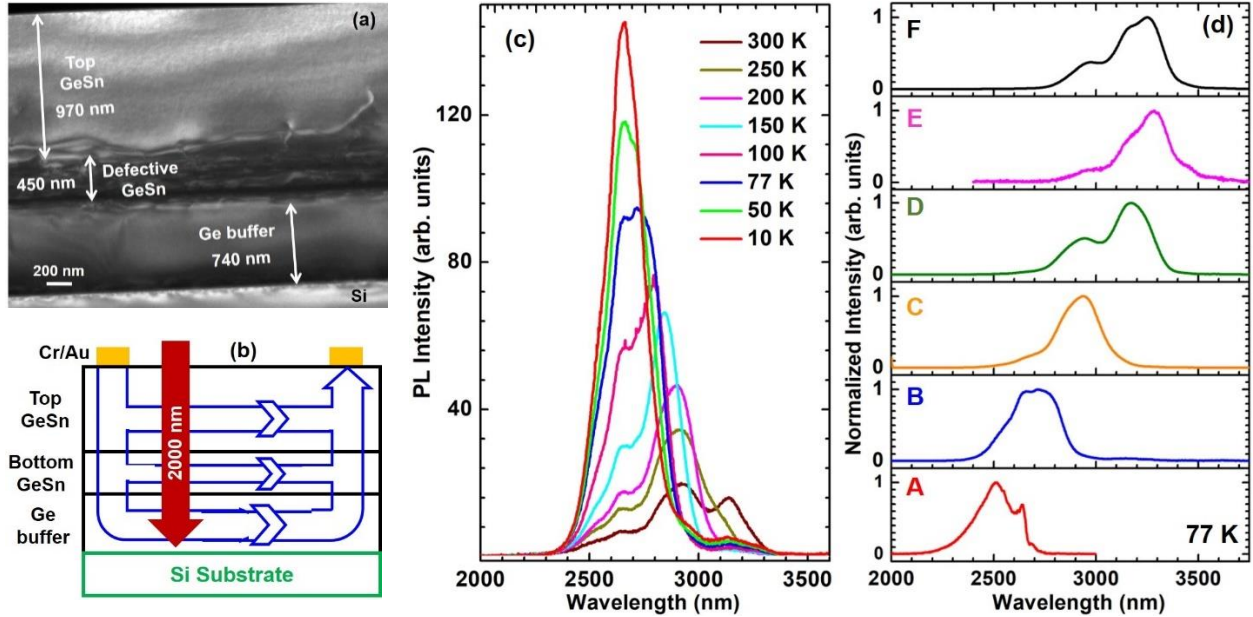

Fig. 1 (a) A typical TEM image of sample E; (b) Schematic drawing of a typical photoconductor, showing penetrating depth of 2 μm laser diode and current flowing (not to scale); (c) Temperature-dependent PL spectra of sample B; (d) PL spectra at 77 K of all samples.

The optical characterization was performed using photoluminescence (PL) technique. For each sample, as the temperature decreases from 300 K to 10 K, the PL intensity dramatically increases. The typical temperature-dependent PL spectra for sample B is shown in Fig. 1 (c). The observed PL peak blue-shift at lower temperature is expected due to the increase of the bandgap.[42] From 300 to 10 K, the integrated PL intensity increases about 12 times. According to band structure calculation, the PL emission is from band-to-band transitions.[43] Note that since the penetration depth of the 532-nm excitation laser beam is less than 100 nm, only the region close to surface involves in the light absorptions. Due to the tilted band edge as a result of compositionally graded Sn content, the photogenerated carriers tend to transport from the wider bandgap region (deeper region in GeSn) to the narrower bandgap region (surface region), i.e., the carrier funneling effect, resulting in light emission mainly occurs at the surface.[43]  The PL spectra of all samples measured at 77 K are shown in Fig. 1 (d).   As the Sn composition increases, the emission peak shifts towards longer wavelength due to the bandgap shrinkage of GeSn as more Sn incorporates.[6, 41]   Note that although sample F has higher Sn composition (22.3%) than that of sample E (20%), the PL peak of sample F was obtained at shorter wavelength, which is attributed to the higher compressive strain of sample F.[41, 43]

**Photoconductor device characterization.**     The GeSn samples were fabricated into photoconductor devices.   Spectral responses of the photoconductors were characterized in the wavelength range from 2.0 µm to 4.0 µm with a Fourier transform infrared (FTIR) spectrometer at a bias of -2 V.  Figure 2 (a) and (b) show the spectral responses of all photoconductors at 300 and 77 K, respectively.   The cut-off wavelength is defined as where the 90% intensity decreases. As the Sn composition increases, the cut-off wavelength shifts towards longer wavelength due to the bandgap shrinkage.[6, 40-43]   At 300 K, the longest cut-off at 3.65 µm was obtained with sample E.

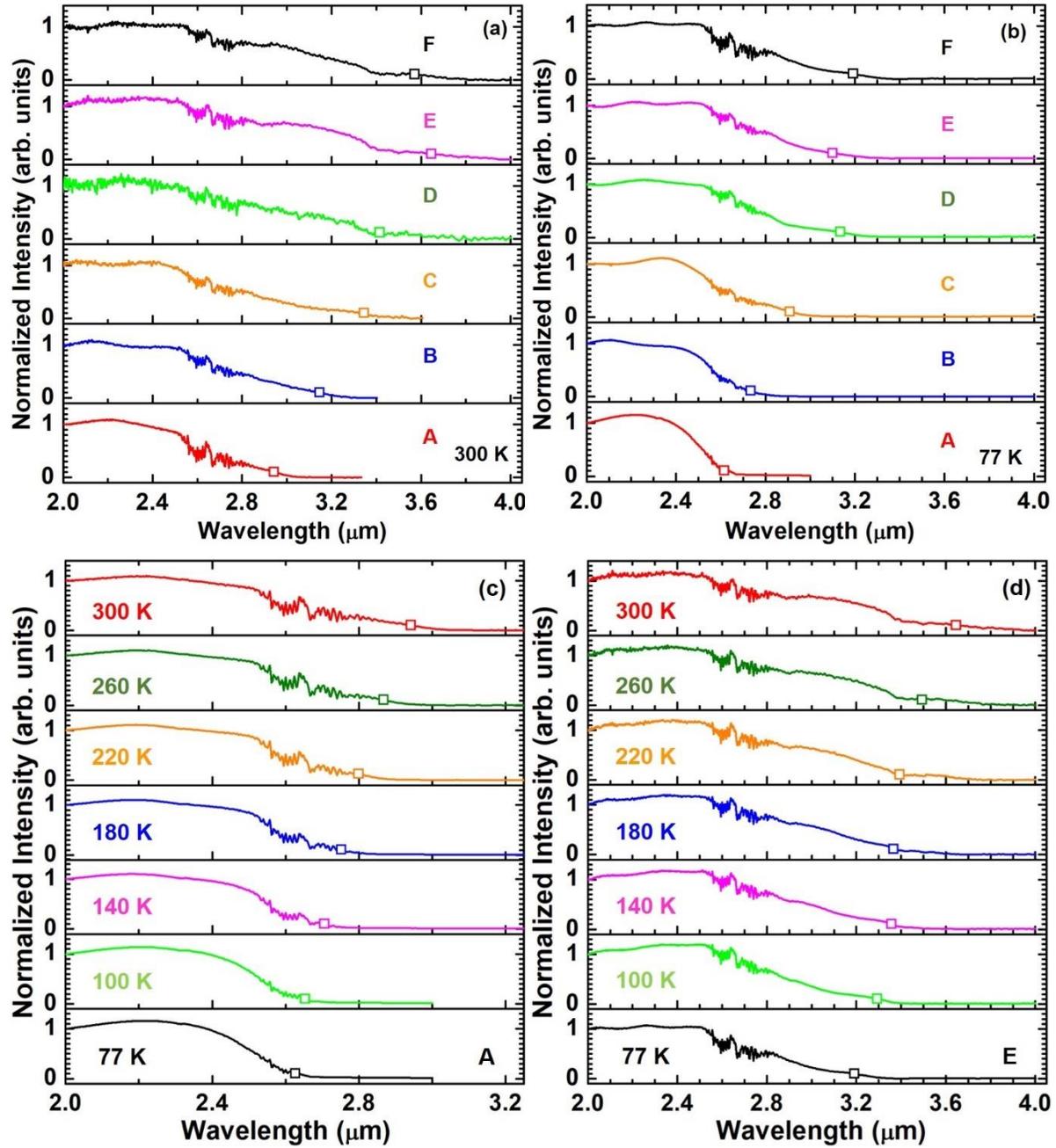

Fig. 2 Spectral response of all devices measured at (a) 300 K and (b) 77 K; Temperature-dependent spectral response of samples (c) A and (d) E (squares represent cut-off wavelengths).

Temperature-dependent spectral response of samples A and E are plotted in Fig. 2 (c) and (d), respectively. As temperature increases, the absorption edge shifts to longer wavelength as expected.[44] From 77 to 300 K, the cut-off wavelengths range from 2.65 to 2.95 µm, and from 3.2 to 3.65 µm for samples A and E, respectively.

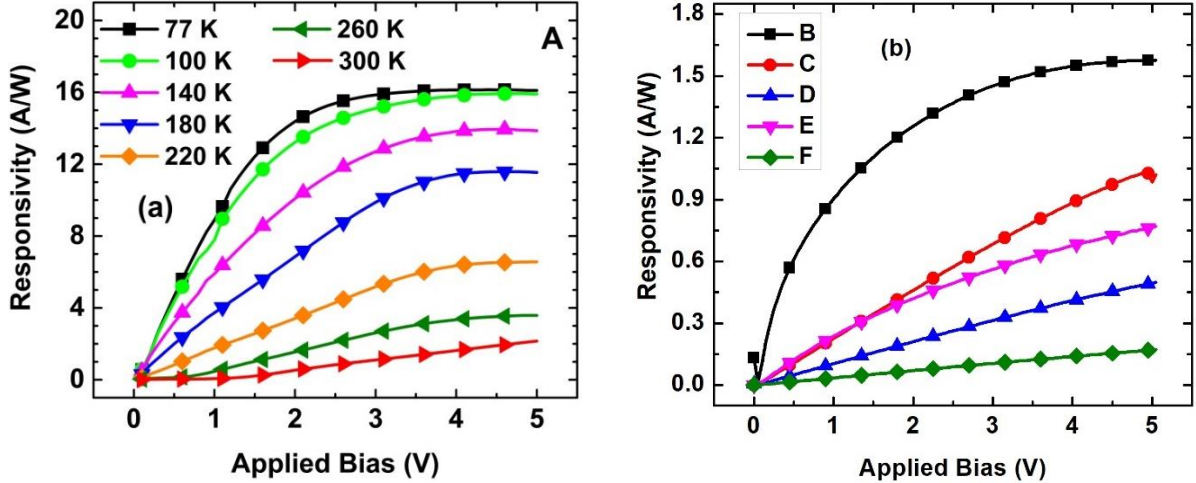

Fig. 3 (a) Temperature-dependent responsivity of sample A; (b) Responsivity of samples B to F at 77 K.

The responsivity was measured using a 2.0-µm laser to eliminate the contribution from Ge buffer (see Fig. 1 (b)). The incident laser beam has a spot size of ~1 mm in diameter and a power density of 8.9 mW/cm$^2$. The temperature-dependent responsivity of sample A is shown in Fig. 3 (a). As temperature decreases, the responsivity increases because of the increases carrier lifetime and mobility at lower temperature. At 77 K, the peak responsivity of 16.1 A/W was obtained at -4 V, which is more than five times higher than our previously reported device (2.85 A/W), which was measured at 1.55 µm) and therefore the Ge absorption also contributes to such device. This can be explained by that the thicker GeSn layer significantly enhances the total light absorptance, resulting in more photo generated carriers. On the other hand, the responsivity increases as the reverse bias voltage increases due to the reduced carrier transit time, which increases the photoconductive gain. However, at each temperature, the responsivity saturation was observed at certain voltage. This is mainly due to Joule heating of the device element and the sweep-out of minority carriers, which limits the applied bias voltage.[27, 28]

Figures 3 (a) and (b) show responsivity of all devices at 77 K. Generally, as the Sn composition increases from samples A to F, the responsivity decreases under the same bias voltage. This can be explained as follows: the higher Sn composition results in narrower bandgap, and consequently the higher intrinsic carrier concentration, which increases the dark current and thus decreases the responsivity.[2, 27] Note that since the quality of samples shows slight difference, the deviation from this trend can be observed, such as samples D and E. The device performance is summarized in Table III.

Table III Summary of photoconductor performance

| Sample | Max. Sn % | Peak D* (cm·Hz$^{1/2}$W$^{-1}$) | Peak Responsivity (A/W) | | Cut-off wavelength (μm) | |
|---|---|---|---|---|---|---|
| | | | 77 K | 300 K | 77 K | 300 K |
| A | 12.5% | $1.1 \times 10^{10}$ | 16.1 | 2.0 | 2.65 | 2.95 |
| B | 15.9% | $5.5 \times 10^{8}$ | 1.6 | $4.4 \times 10^{-2}$ | 3.00 | 3.20 |
| C | 15.7% | $6.2 \times 10^{8}$ | 1.0 | $7.2 \times 10^{-3}$ | 3.15 | 3.40 |
| D | 17.9% | $4.4 \times 10^{8}$ | 0.5 | $3.8 \times 10^{-3}$ | 3.30 | 3.35 |
| E | 20.0% | $1.1 \times 10^{8}$ | 0.8 | $6.7 \times 10^{-3}$ | 3.00 | 3.65 |
| F | 22.3% | $1.1 \times 10^{8}$ | 0.2 | $3.2 \times 10^{-3}$ | 3.35 | 3.65 |

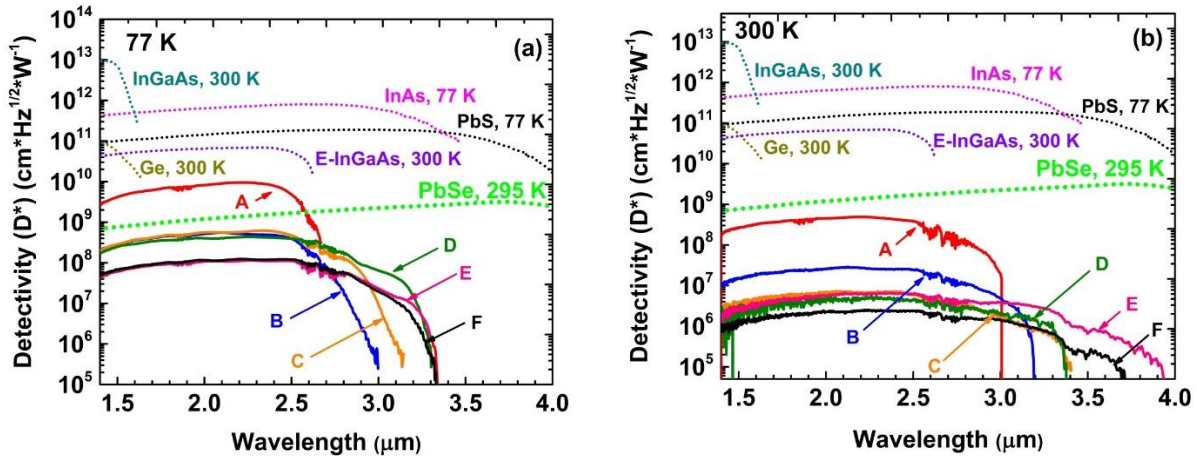

Fig. 4 Spectral D* of all devices at (a) 77 K and (b) 300 K.

Specific detectivities (D*) were determined using noise equivalent power (NEP) at 1Hz frequency bandwidth and device area, where the NEP was calculated based on mean squared noise current and responsivity. The spectral D* of all devices at 77 and 300 K are plotted in Fig. 4 to compare with commercial photodetectors. The D* dropped as higher temperature, which is mainly attributed to the increased noise at higher temperatures.[27, 28, 35, 40] Among all devices, sample A exhibits highest peak D* due to its relative high responsivity. At 2.4 μm, the peak D* of $1.1 \times 10^{10}$ cm·Hz$^{1/2}$W$^{-1}$ was achieved. Note that at the wavelength range from 1.5 to 2.6 μm, the D* of sample A exceeds that of commercial PbSe detector, and only a few times smaller than extended-InGaAs and PbS detectors. The D* and cut-off wavelength of each device are summarized in Table III.

**Photodiode device characterization**. Our previous work on the photodiode detector indicated that the device performance can be dramatically improved by passivation of the surface and sidewall to reduce leakage currents.[35] In this work, a Ge$_{0.89}$Sn$_{0.11}$ photodiode was grown and fabricated (500 μm of mesa diameter) to develop the passivation technique. Note that although photoconductors show higher responsivity and D*, passivation could not further improve their performance since surface leakage current is not dominant in photoconductor, but the current leakage in the Ge buffer. Therefore, we selected photodiode for passivation study.

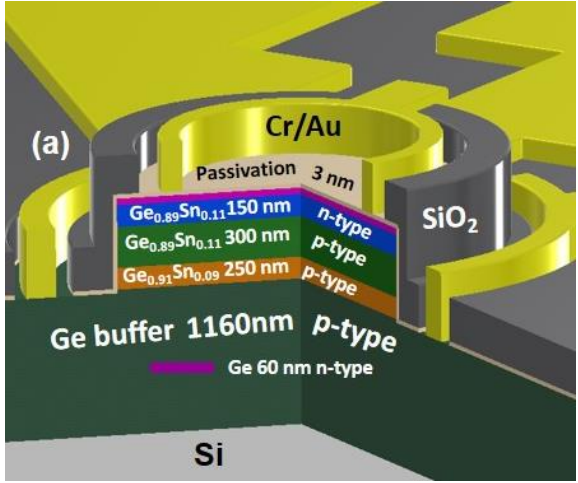
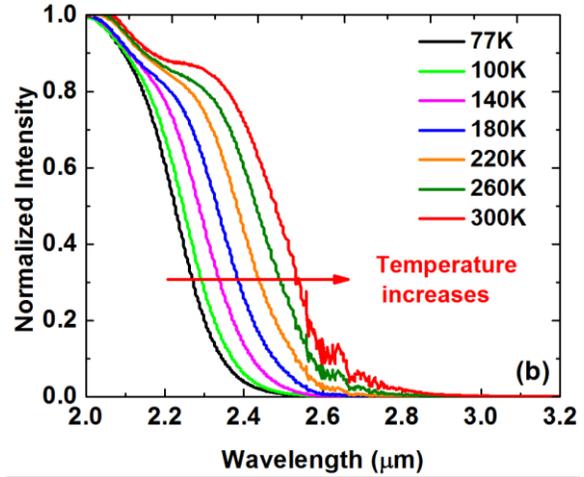

Fig. 5 (a) Device structure schematic diagram; (b) Temperature-dependent spectral response.

Figure 5 (a) shows the cross-sectional view of a schematic diagram for the photodiode. The GeSn shows two-layer feature, with a 250-nm-thick defective bottom layer and a 450-nm-thick high-quality top layer. The n- and p-type doping concentrations of the $Ge_{0.89}Sn_{0.11}$ layer were measured as $2 \times 10^{18}$ cm$^{-3}$ and $1 \times 10^{17}$ cm$^{-3}$, respectively. As a result, the depletion region (calculated as 150-nm width) is located mainly in the p-type GeSn layer with high material quality. Since the depletion region edge is ~150 nm away from the defective GeSn bottom layer, most photo generated carriers can be effectively collected. For the passivated device, a 3-nm-thick GeON was deposited, which covers the top and sidewall, as shown in Fig. 5 (a).

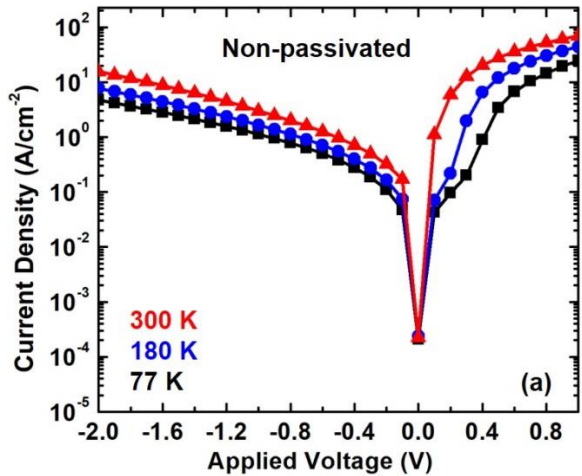
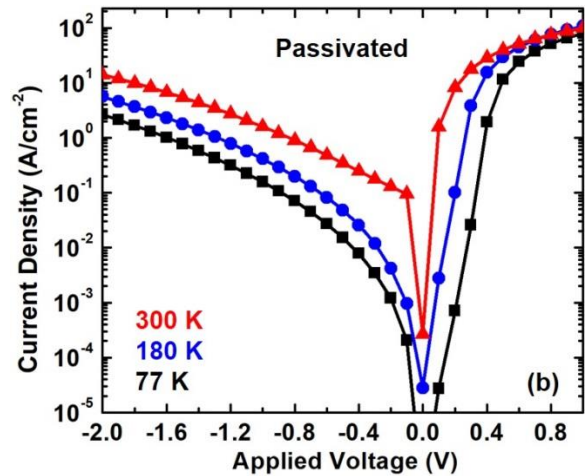

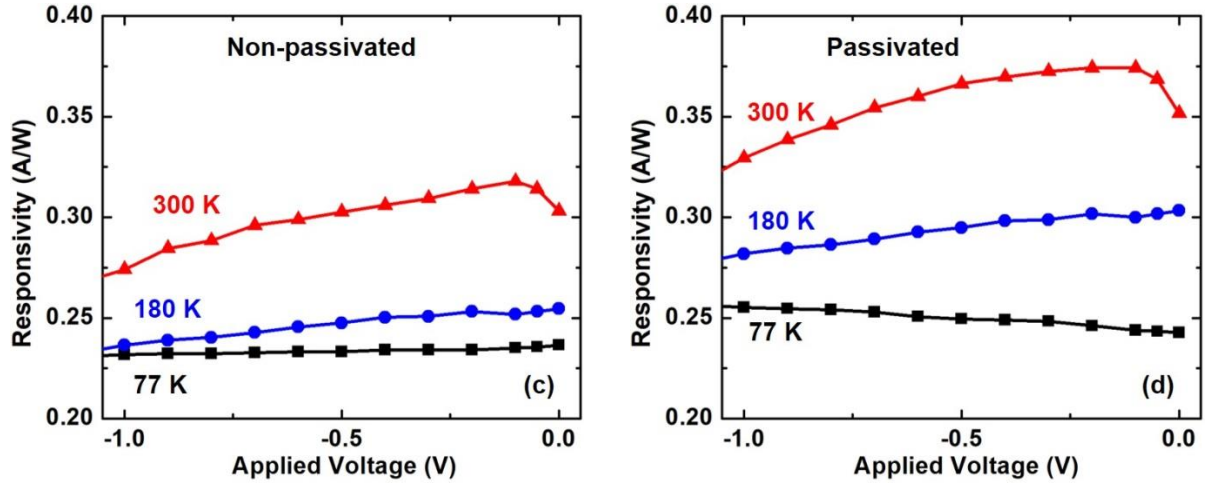

Fig. 6 Dark current-voltage characteristics of (a) non-passivated and (b) passivated device; responsivity for (c) non-passivated and (d) passivated devices.

The temperature-dependent spectral response for the passivated device is shown in Fig. 5 (b). As temperature increases from 77 to 300 K, the cut-off wavelength shifts towards long wavelength, from 2.3 to 2.7 μm, which is similar to the non-passivated device as expected.

Dark current-voltage characteristics were measured with a direct current (DC) source. Figure 6 (a) and (b) show the current-voltage characteristics for devices without and with passivation from 300 to 77 K, respectively. The rectifying diode-like behavior was observed for each curve. Compared with the non-passivated device, the passivated device exhibits ~50% lower reverse dark current under lower bias at all temperatures. To better understand the effect of passivation, an in-depth study on mechanism of each dark current component and its comparison between non-passivated and passivated devices was conducted via data fitting (see supplementary), which explains the reduced dark current as following: passivation reduces the surface defect density of states, trap density, and thus generation-recombination centers. Therefore, the shunt resistance, trap-assisted tunneling resistance, and generation-recombination resistance increase, resulting in reduced corresponding current components.

Figure 6 (c) and (d) shows the comparison of responsivity between non-passivated and passivated devices. For each device, the responsivity decreases as the temperature decreases. This is due to: i) the bandgap increases at lower temperature, which results in reduced light absorption; and ii) trap states in high-quality GeSn layer affect the non-radiative recombination of the photo-generated carriers. At lower temperature, the photo-excited carriers partially recombine at the deep level trap, where they cannot gain sufficient thermal energy and then to be released; while at higher temperatures, the carriers could recombine through the shallow trap level, where they can be released by thermal excitation, leading to increased responsivity at higher temperature. For the non-passivated device, the peak responsivity of 0.32 A/W was obtained at 300 K. For the passivated device, due to reduced dark current, the peak responsivity of 0.38 A/W was achieved at 300 K, showing ~20% improvement.

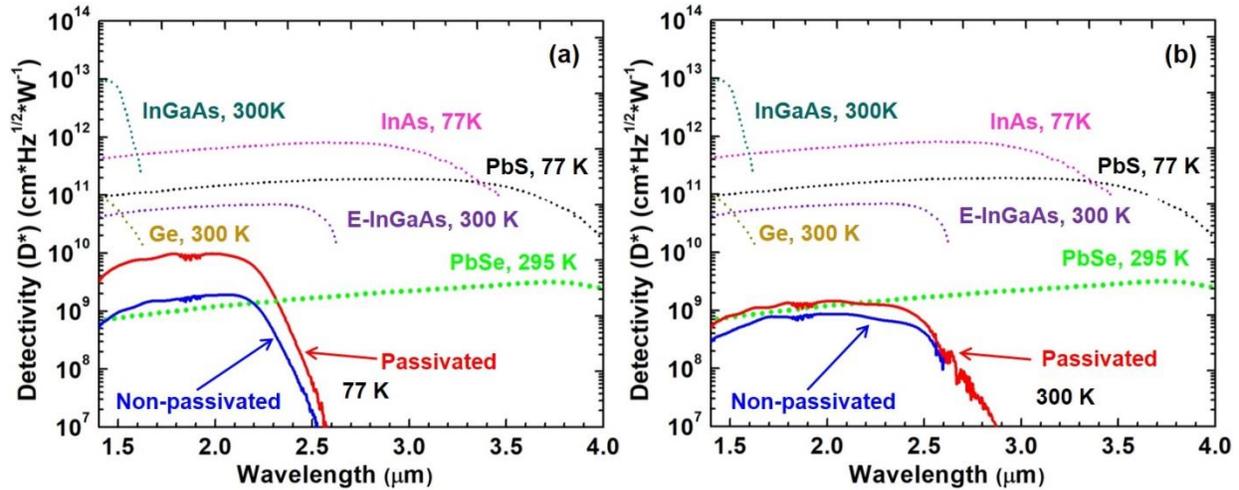

Fig. 7 D* of non-passivated and passivated devices at (a) 77 K and (b) 300 K.

The spectral D* at 77 and 300 K are plotted in Fig. 7 (a) and (b), respectively. At 77 K, the peak D* of $1.1 \times 10^{10}$ cm·Hz$^{1/2}$·W$^{-1}$ was obtained for the passivated device, which is about four times higher than that of the non-passivated device ($2.0 \times 10^9$ cm·Hz$^{1/2}$·W$^{-1}$). It is worth noting that this D* value is superior to PbSe detector at the given wavelength range and is comparable to that of the commercial extended-InGaAs detector (~4×10$^{10}$ cm·Hz$^{1/2}$·W$^{-1}$) at the same wavelength range. Even at 300 K, the passivated device shows a little better D* compared to the PbSe detector from 1.5 to 2.2 µm.

**Demonstration of mid-infrared imaging.** The mid-infrared imaging using discrete photoconductor and discrete photodiode were further demonstrated. The major components of imaging setup include white light source covering 300 to 2400 nm, focus lenses, 3D stage for scanning, and long-pass filters to allow infrared light pass. The setup and experimental method are detailed in the supplementary.

| Object | Photoconductors at 77 K | | Commercial PbSe Devices at 300 K | |
|---|---|---|---|---|
| 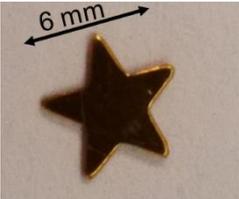 | 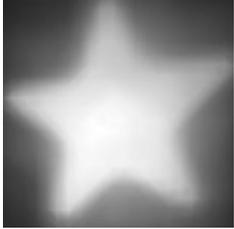 | 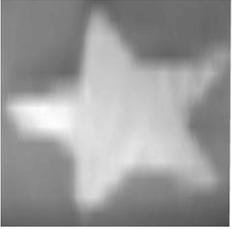 | 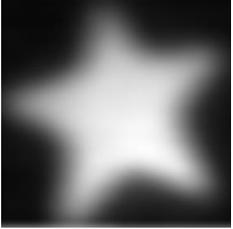 | 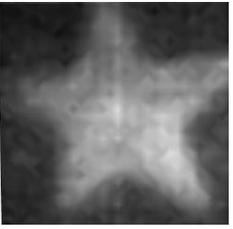 |
| | Device A | Device E | PbSe Single Detector | PbSe Camera |

| Passivated Photodiode | | More infrared imaging results at 300 K | | |
|---|---|---|---|---|
| 77 K | 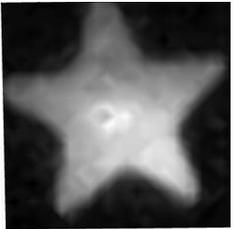 | Objects | 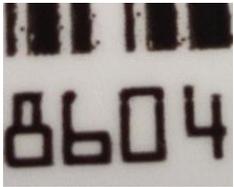 | 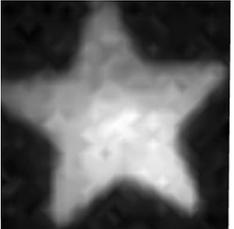 |
| 300 K | 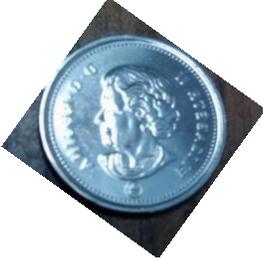 | Passivated Photodiode | 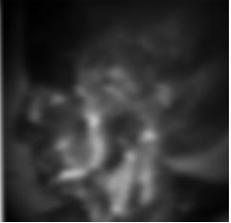 | 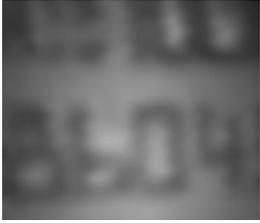 |

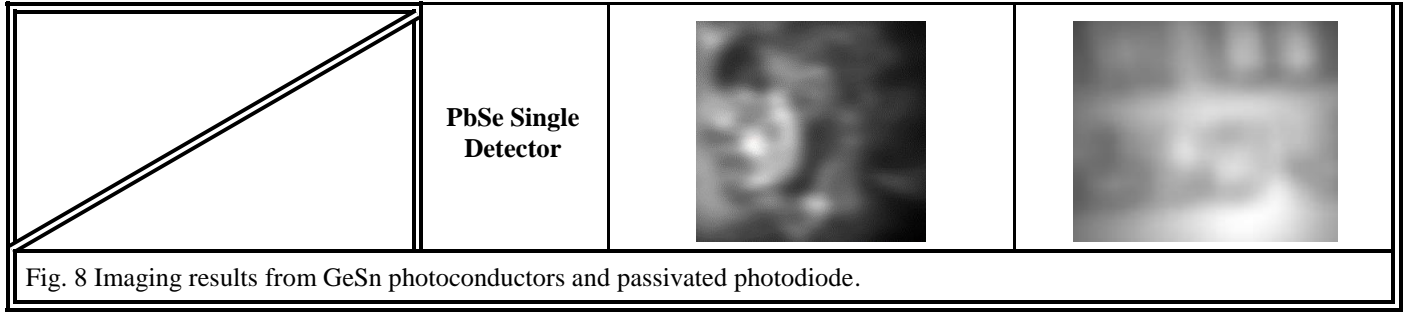

Fig. 8 Imaging results from GeSn photoconductors and passivated photodiode.

For photoconductors, all devices could demonstrate the IR imaging at 77 K with standard lock-in technique to reduce the noise. It is worth noting that for devices A and E, the images can be clearly obtained even without lock-in amplifier, as shown in Fig. 8 top. The quality of image is better than that of commercial PbSe single detector and camera (32 by 32 pixels). For photodiodes, both non-passivated and passivated devices could demonstrate the IR imaging at 77 and 300 K. The images taken from passivated device at 77 and 300 K are shown in Fig. 8 bottom-left. The quality of both images is comparable with PbSe detectors, with the image at 77 K being better than that at 300 K. Some other objects including a coin and a barcode were used for more imaging demonstration using the passivated photodiode and the PbSe detector at 300 K, as shown in Fig. 8 bottom-right. Compared with PbSe detector, the images taken from GeSn photodiode exhibit clearer features.

In conclusion, high-performance GeSn photodetectors were presented in this work. We demonstrated a set of photoconductors with Sn compositions up to 22.3%. The longest wavelength cut-off was measured up to 3.65 µm. The peak D* is comparable with commercial extended InGaAs detector. The passivation technique was developed and applied on a $Ge_{0.89}Sn_{0.11}$ photodiode. The significantly enhanced responsivity spectral D* were achieved. Moreover, the mid-IR imaging was demonstrated with GeSn photoconductors and photodiode. The images taken from GeSn photodetectors showed better quality compared with commercial PbSe detector.

## Methods

**Material growth and characterization.** The GeSn samples were grown using an industry standard ASM Epsilon® 2000 Plus reduced pressure chemical vapor deposition system with $SnCl_4$ and $GeH_4$ as Sn and Ge precursors, respectively. A thick relaxed Ge buffer was first grown on the Si substrate as buffer layer. Then a thick GeSn layer was grown with a multiple-step Sn-enhanced growth recipe. The details of the growth technique were published elsewhere.[9,10] After growth, material characterizations including cross-sectional TEM and secondary ion mass spectrometry (SIMS) were performed to identify the layer thicknesses and Sn composition. The Sn composition was cross checked by data fitting of reciprocal space mapping (RSM), and the strain information was also obtained by RSM.

The temperature-dependent PL measurements were performed using a standard off-axis configuration with a lock-in technique (optically chopped at 377 Hz). A 532-nm continuous wave (CW) laser with spot size of 65 µm and power density of 15 $kW/cm^2$ was used as excitation source. The emissions were collected using a spectrometer equipped with an InSb detector with cut-off wavelength at 5.0 µm.

**Device Fabrication.** The GeSn samples were fabricated into photoconductors by photolithography and wet chemical etch processes. The mesa structures with areas of $500 \times 500$ $\mu m^2$ were etched by using the solution of HCl: $H_2O_2$: $H_2O$ = 1:1:10 at 0 °C. The etching depth was controlled to reach Si substrate. A 300-nm-thick $SiO_2$ passivation layer was then deposited by electron beam evaporator followed by the openings made for the metal contacts. Electrode pads were patterned and metalized with 10/300 nm of Cr/Au. The interdigitated electrodes with 12 μm finger width and 24 μm spacing were used in order to enhance the gain of the photoconductors. The decreasing electrode spacing was expected to increase the photoconductive gain by reduction of the carrier transit time.

**Data availability**

The data that support the findings of this study are available from the corresponding author upon reasonable request.

## Acknowledgements

The work was supported by the National Aeronautics and Space Administration Established Program to Stimulate Competitive Research (NASA EPSCoR) (Grant No. NNX15AN18A) and the Air Force Office of Scientific Research (AFOSR) (Grant Nos. FA9550-14-1-0205, FA9550-16-C-0016). Dr. Wei Du appreciates support from Provost's Research & Scholarship Fund at Wilkes University. We are thankful for Dr. M. Benamara's assistance in TEM imaging and Dr. A. Kuchuk's assistance in XRD measurement from Institute for Nanoscience & Engineering, University of Arkansas. We also thank Dr. Tansel Karabacak and Emad Badradeen from Department of Physics and Astronomy in University of Arkansas at Little Rock for their effort in passivation work.

## Author Contributions

H.T. and T.P. contributed equally to this work. S.Y., M.M., B.L., and W.Du proposed and guided the overall project. J.M. and J.T. planned and conducted GeSn epitaxial growths. J.M., W.Dou, P.G. performed material characterization. Y.Zhou, P.G., and S.A. conducted PL measurements. H.T., T.P., and Y.Zhou fabricated photodetector devices. J.G. developed device structure schematic diagrams. H.T. and T.P. developed, conducted, and analyzed photodetector measurements. Y.Zhang performed device physics analysis. All authors discussed the results and commented on the manuscript.

## Additional information

Supplementary Information accompanies this paper at xxxxx.

## Competing Interests

The authors declare no competing interests.

**Corresponding author**

Correspondence to Shui-Qing Yu.